\input{aipcheck}

\documentclass[
    ,final            
  ]
  {aipproc}

\layoutstyle{6x9}

\begin{document}

\title{The strong coupling constant at large distances}


\author{A. Deur}{
  address={Thomas Jefferson National Accelerator Facility, 
Newport News, VA 23606}}

\begin{abstract}
In this paper we discuss effective strong coupling constants. 
Those are well behaved in the low-$Q^2$ 
domain, contrarily to $\alpha_s$ from pQCD. We present an 
extraction of an effective strong coupling constant from Jefferson 
Lab polarized data at intermediate and low $Q^2$. We also 
show how these data, together with spin sum rules, allow us to 
obtain the effective coupling constant 
over the entire $Q^2$ range. We then discuss the relation 
between the experimentally extracted coupling 
constant and theoretical calculations at low $Q^2$. We 
conclude on the importance of such study for the 
application of the AdS/CFT correspondence to QCD.
\end{abstract}

\maketitle


In QCD, the magnitude of the strong force is given by 
the running coupling constant $\alpha_{s}$. 
At large $Q^2$, in the pQCD domain, $\alpha_{s}$ is well defined and 
is given by the series:
\begin{eqnarray}
\mu\frac{\partial\alpha_{s}}{\partial\mu} & =2\beta(\alpha_{s}) & 
=-\frac{\beta_{0}}{2\pi}\alpha_{s}^{2}-\frac{\beta_{1}}{4\pi^{2}}
\alpha_{s}^{3}-\frac{\beta_{2}}{64\pi^{3}}\alpha_{s}^{4}-...
\label{eq:alpha_s beta serie}
\end{eqnarray}
Where $\mu$ is the energy scale, to be identified to $Q$. The 
first terms of the $\beta$ series are: 
$\beta_{0}=11-\frac{2}{3}n$ with $n$ the number of active quark 
flavors, $\beta_{1}=51-\frac{19}{3}n$ and $\beta_{2}=2857-
\frac{5033}{9}n+\frac{325}{27}n^{2}$.
The solution of the differential equation \ref{eq:alpha_s beta serie} is:
\small
\begin{eqnarray}
\alpha_{s}(\mu)=\frac{4\pi}{\beta_{0}ln(\mu^{2}/\Lambda_{QCD}^{2})}
\times \label{eq:alpha_s}\\
 &  & \hspace{-6cm} \left[1-\frac{2\beta_{1}}{\beta_{0}^{2}}
\frac{ln\left[ln(\mu^{2}/\Lambda_{QCD}^{2})\right]}{ln(\mu^{2}/
\Lambda_{QCD}^{2})}+\frac{4\beta_{1}^{2}}{\beta_{0}^{4}ln^{2}
(\mu^{2}/\Lambda_{QCD}^{2})}\left(\left(ln\left[ln(\mu^{2}/
\Lambda_{QCD}^{2})\right]-\frac{1}{2}\right)^{2}+\frac{
\beta_{2}\beta_{0}}{8\beta_{1}^{2}}-\frac{5}{4}\right)\right]
\nonumber 
\end{eqnarray}
\normalsize

\noindent Eq. \ref{eq:alpha_s} allows us to evolve the different experimental 
determinations of $\alpha_{s}$ to a conventional scale, typically 
$M_{z_{0}}^{2}$. 
The agreement between the $\alpha_{s}$ obtained from different observables
demonstrates its universality and the validity of 
Eq. \ref{eq:alpha_s beta serie}. One can obtain  
$\alpha_{s}(M_{z_{0}}^{2})$ with
doubly polarized DIS data and assuming the validity of the Bjorken sum 
rule~\cite{Bjorken}: 
\begin{equation}
\Gamma_{1}^{p-n}=\int_{0}^{1}(g_{1}^{p}-g_{1}^{n})dx=\frac{g_{a}}{6}[1-\frac{\alpha_{{\rm {s}}}}{\pi}-3.58\left(\frac{\alpha_{{\rm {s}}}}{\pi}\right)^{2}-20.21\left(\frac{\alpha_{{\rm {s}}}}{\pi}\right)^{3}+...]+O(\frac{1}{Q^{2}}),\label{eq:genBj2}
\end{equation}
where $g_A$ is the well measured nucleon axial charge. Solving 
Eq. \ref{eq:genBj2} using the experimental value of 
$\Gamma_{1}^{p-n}$, and then using Eq. \ref{eq:alpha_s} provides 
$\alpha_{s}(M_{z_{0}}^{2})$.

Eq. \ref{eq:alpha_s} leads to an infinite coupling at large distances, 
when $Q^2$ approaches $\Lambda^{2}_{QCD}$, 
This is not a conceptual problem since we are out of the validity 
domain of pQCD on which Eq. \ref{eq:alpha_s} is based.
But since data show no sign of discontinuity or phase transition 
when crossing the intermediate $Q^{2}$ domain, one 
should be able to define an effective coupling $\alpha_{s}^{eff}$ at 
any $Q^2$ that matches $\alpha_{s}$ at large $Q^{2}$ but stays finite at 
small $Q^{2}$. 

The Bjorken Sum Rule can be used again to define $\alpha_{s}^{eff}$ at 
low Q$^{2}$. Defining $\alpha_{s}^{eff}$
from Eq. (\ref{eq:genBj2}) truncated to first order: 
$\Gamma_{1}^{p-n}\equiv\frac{1}{6}(1-\alpha_{s,g_{1}}/\pi)$,
offers many advantages. In particular, $\alpha_{s}^{eff}$ does not 
diverge near $\Lambda_{QCD}$ and is renormalization scheme independent 
since the first term in a pQCD series is the same, regardless to the choice
of renormalization scheme. However, $\alpha_{s}^{eff}$ becomes dependent 
on the choice of observable employed to define it. If $\Gamma_{1}^{p-n}$ is 
used as the defining observable, the effective coupling is noted 
$\alpha_{s,g_{1}}$.  
Relations, called \emph{commensurate scale relations} \cite{Brodsky CSR}, 
link the different effective couplings so in principle one
effective coupling is enough to describe the strong force and the theory 
retains its predictive power.

\begin{figure}
\includegraphics[scale=0.26]{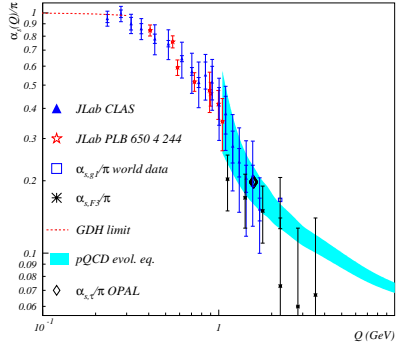}
\caption{\label{fig: alpha_s eff} 
Value of $\alpha_{s,g_{1}}/\pi$ 
extracted from the world data on the Bjorken sum at 
$Q^{2}=5$ GeV$^{2}$ \cite{SLAC} and from recent JLab data \cite{JLabAB}
used to extract the Bjorken Sum \cite{JLab}. Also shown are 
$\alpha_{s,\tau}$ extracted from the OPAL 
data on $\tau$ decay \cite{Brodsky CSR}, and $\alpha_{s,GLS}$ extracted 
using the 
Gross-Llewellyn Smith sum rule \cite{GLS} and its measurement by the 
CCFR collaboration \cite{CCFR}.
The values of $\alpha_{s,g_{1}}$ extracted using the pQCD expression 
of the Bjorken sum at leading twist and third order in $\alpha_{s}$ (with $\alpha_{s}$ 
computed using Eq. \ref{eq:alpha_s}) is given by the gray band. The values of 
$\alpha_{s,g_{1}}/\pi$ extracted using the GDH 
sum rule is given by the red dashed line.} 
\end{figure}

The effective coupling definition in term of 
pQCD evolution equations truncated to first order was proposed by 
Grunberg \cite{Grunberg}. Following this 
definition, effective couplings have been extracted from different 
observables and have been compared to each other
using the commensurate scale relations \cite{deur alpha_s^eff}, see 
Fig. \ref{fig: alpha_s eff}.
There is good agreement between the effective couplings 
$\alpha_{s,g_{1}}$, $\alpha_{s,F_{3}}$ and
 $\alpha_{s,\tau}$. The GDH and Bjorken sum rules can be used to extract  
$\alpha_{s,g_{1}}$ at small and large $Q^{2}$ respectively 
\cite{deur alpha_s^eff}. This, together with the JLab data at 
intermediate $Q^{2}$, provides for the first time a coupling 
at any $Q^{2}$. A striking feature of Fig. \ref{fig: alpha_s eff} 
is that $\alpha_{s,g_{1}}$ becomes scale invariant at small $Q^{2}$. 
This was predicted by a number of calculations and it is 
known that color confinement leads to an
infrared fixed point \cite{irfp}, but it is the 
first time it is seen experimentally.

\begin{figure}
\vspace{-0.8cm}
\includegraphics[scale=0.37]{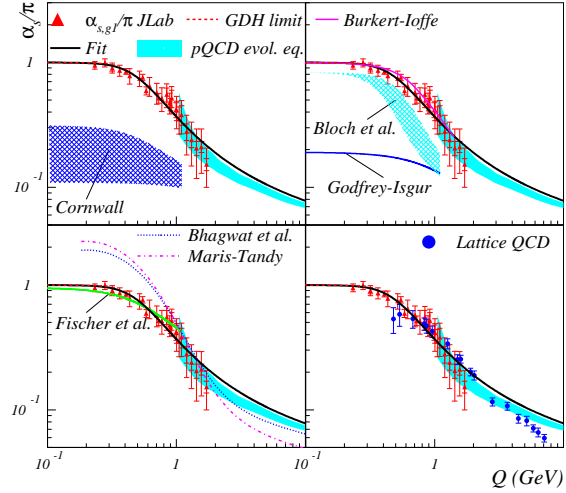}
\vspace{-0.7cm}
\caption{\label{fig: alpha_s eff4} The effective coupling 
$\alpha_{s,g_{1}}$ extracted from JLab data, its fit, and its 
extraction using the Burkert and Ioffe \cite{bur92} model to 
obtain $\Gamma_{1}^{p-n}$. The $\alpha_{s}$ calculations are: 
Top left: Schwinger-Dyson equations (Cornwall \cite{Cornwall}); 
Top right: Schwinger-Dyson equations (Bloch) \cite{Bloch} and 
$\alpha_{s}$ used in a quark constituent model \cite{Godfrey-Isgur};
Bottom left: Schwinger-Dyson equations (Maris-Tandy \cite{Tandy}),
Fischer, Alkofer, Reinhardt and Von Smekal \cite{Fischer} and Bhagwat 
et al. \cite{Bhagwat}; Bottom right: Lattice QCD \cite{Furui}.
}
\end{figure}

A fit of the $\alpha_{s,g_{1}}$ data and sum rule constraints has been 
performed with a form based on Eq. \ref{eq:alpha_s} at 
first order:
\begin{equation}
\alpha^{fit}_{s,g_1}=\gamma n(Q)/log(\frac{Q^{2}+m_{g}^{2}(Q)}{\Lambda^2})
\label{eq: fit}
\end{equation}
\noindent 
where $\gamma=4/\beta_{0}=12/(33-8)$, 
$n=\pi(1+[\frac{\gamma}{log(m^2/\Lambda^2)(1+Q/\Lambda)-\gamma}+(bQ)^{c}]^{-1})$ and 
$m_{g}=(m/(1+(aQ)^{d}))$. The values of the parameters are:
$\Lambda=0.349 \pm 0.009$ GeV,
$a=3.008 \pm 0.081$ GeV$^{-1}$,
$b=1.425 \pm 0.032$ GeV$^{-1}$,
$c=0.908 \pm 0.025$,
$m=1.204  \pm 0.018$ GeV,
$d=0.840  \pm 0.051$. 
$m_{g}$ has been  interpreted as an effective gluon 
mass \cite{Cornwall}. The fit is shown on Fig. \ref{fig: alpha_s eff4} 
(continuous black line). Eq. \ref{eq: fit}, used in 
$\Gamma_{1}^{p-n}\equiv\frac{1}{6}(1-\alpha_{s,g_{1}}/\pi)$, can 
also be employed to parametrize the generalized Bjorken and GDH sums.

On Fig. \ref{fig: alpha_s eff4}, $\alpha_{s,g_{1}}$ is compared to 
theoretical results. There are several techniques used to predict 
$\alpha_{s}$ at small $Q^{2}$, e.g. lattice QCD, solving the 
Schwinger-Dyson equations, or choosing the coupling in a constituent 
quark model so that it reproduces hadron spectroscopy. However, the 
connection 
between  these $\alpha_{s}$ is unclear, in part because of the different 
approximations used. In addition, the precise relation 
between $\alpha_{s,g_{1}}$ (or any effective coupling defined using 
\cite{Grunberg} or \cite{Brodsky CSR}) and these computations is
unknown. Nevertheless, one can still compare them to see 
if they share common features. The calculations and $\alpha_{s,g_{1}}$ 
present 
a similar behavior. Some calculations, in
particular the lattice one, are in excellent agreement with 
$\alpha_{s,g_{1}}$. 

These works show that $\alpha_{s}$ is scale invariant (\emph{conformal 
behavior}) at small 
and large $Q^{2}$ (but not in the transition region between the 
fundamental description of QCD in terms of quarks and gluons 
degrees of freedom and its effective one in terms of baryons and mesons).
The scale invariance at large $Q^2$ is the well known asymptotic freedom. 
The conformal behavior at small $Q^{2}$ 
is essential to apply a property of \emph{conformal field theories} 
(CFT) to the study of hadrons: the \emph{Anti-de-Sitter 
space/Conformal Field Theory (AdS/CFT) correspondence} of Maldacena 
\cite{Maldacena}, that links a strongly coupled gauge field to weakly 
coupled superstrings states. Perturbative calculations are feasible in 
the weak coupling AdS theory. They are then projected on the AdS 
boundary, where they correspond to the calculations that would have
been obtained with the strongly coupled CFT. This opens the possibility 
of analytic non-perturbative QCD calculations \cite{ads/CFT}.

To sum up, thanks to the data on nucleon spin structure and to spin 
sum rules, an effective strong coupling can be extracted in any regime
of QCD. The question of comparing it with theoretical calculations of
$\alpha_{s}$ at low $Q^2$ is open, but such comparison exposes a
similarity between these couplings. Apart for the parton-hadron transition 
region, the coupling shows
that QCD is approximately a conformal theory. This is a necessary
ingredient to the application of the AdS/CFT correspondence that may 
make analytical calculations possible in the non-perturbative domain of QCD.

\textbf{Acknowledgments} This work was done in collaboration with 
V. Burkert, J-P. Chen and W. Korsch. It is supported by the U.S. 
Department of Energy (DOE). The Jefferson Science Associates (JSA) 
operates the Thomas Jefferson National Accelerator Facility for the DOE 
under contract DE-AC05-84ER40150.


\IfFileExists{\jobname.bbl}{}
 {\typeout{}
  \typeout{******************************************}
  \typeout{** Please run "bibtex \jobname" to optain}
  \typeout{** the bibliography and then re-run LaTeX}
  \typeout{** twice to fix the references!}
  \typeout{******************************************}
  \typeout{}
 }

\end{document}